\pdfoutput=1
\documentclass[12pt,a4paper]{article}
\usepackage{ifthen}
\newboolean{pdflatex}
\setboolean{pdflatex}{true}

\newboolean{articletitles}
\setboolean{articletitles}{true}

\newboolean{uprightparticles}
\setboolean{uprightparticles}{false}

\newboolean{inbibliography}
\setboolean{inbibliography}{false}

\usepackage[top=1in, bottom=1.25in, left=1in, right=1in]{geometry}

\usepackage{microtype}
\usepackage{lineno}  
\usepackage{xspace} 
                    
\usepackage{caption}

\usepackage{graphicx}  
\usepackage{color}
\usepackage{colortbl}
\graphicspath{{./figs/}} 

\usepackage{amsmath} 
\usepackage{amssymb}
\usepackage{amsfonts}
\usepackage{upgreek} 

\newcommand*\patchAmsMathEnvironmentForLineno[1]{%
\expandafter\let\csname old#1\expandafter\endcsname\csname #1\endcsname
\expandafter\let\csname oldend#1\expandafter\endcsname\csname
end#1\endcsname
 \renewenvironment{#1}%
   {\linenomath\csname old#1\endcsname}%
   {\csname oldend#1\endcsname\endlinenomath}%
}
\newcommand*\patchBothAmsMathEnvironmentsForLineno[1]{%
  \patchAmsMathEnvironmentForLineno{#1}%
  \patchAmsMathEnvironmentForLineno{#1*}%
}
\AtBeginDocument{%
\patchBothAmsMathEnvironmentsForLineno{equation}%
\patchBothAmsMathEnvironmentsForLineno{align}%
\patchBothAmsMathEnvironmentsForLineno{flalign}%
\patchBothAmsMathEnvironmentsForLineno{alignat}%
\patchBothAmsMathEnvironmentsForLineno{gather}%
\patchBothAmsMathEnvironmentsForLineno{multline}%
\patchBothAmsMathEnvironmentsForLineno{eqnarray}%
}

\usepackage{hyperref}    
\usepackage[all]{hypcap} 

\usepackage{xspace} 
\usepackage{upgreek}

\def\lhcb {\mbox{LHCb}\xspace}

\def\rich   {RICH\xspace}

\def\MagUp {\mbox{\em Mag\kern -0.05em Up}\xspace}

\ifthenelse{\boolean{uprightparticles}}%
{

 \def\Pmu         {\ensuremath{\upmu}\xspace}

 \def\Ppi         {\ensuremath{\uppi}\xspace}

 \def\Ppsi        {\ensuremath{\uppsi}\xspace}

 \def\PDelta      {\ensuremath{\Delta}\xspace}                 
 \def\PXi      {\ensuremath{\Xi}\xspace}                 
 \def\PLambda      {\ensuremath{\Lambda}\xspace}                 
 \def\PSigma      {\ensuremath{\Sigma}\xspace}                 
 \def\POmega      {\ensuremath{\Omega}\xspace}                 
 \def\PUpsilon      {\ensuremath{\Upsilon}\xspace}

 \def\PB      {\ensuremath{\mathrm{B}}\xspace}                 
                  
 \def\PD      {\ensuremath{\mathrm{D}}\xspace}

 \def\PJ      {\ensuremath{\mathrm{J}}\xspace}                 
 \def\PK      {\ensuremath{\mathrm{K}}\xspace}

 \def\Pb      {\ensuremath{\mathrm{b}}\xspace}                 
 \def\Pc      {\ensuremath{\mathrm{c}}\xspace}                 
 \def\Pd      {\ensuremath{\mathrm{d}}\xspace}

 \def\Pi      {\ensuremath{\mathrm{i}}\xspace}

 \def\Ps      {\ensuremath{\mathrm{s}}\xspace}                 
                  
 \def\Pu      {\ensuremath{\mathrm{u}}\xspace}

}
{

 \def\Pmu         {\ensuremath{\mu}\xspace}

 \def\Ppi         {\ensuremath{\pi}\xspace}

 \def\Ppsi        {\ensuremath{\psi}\xspace}                 
                  
 \mathchardef\PDelta="7101
 \mathchardef\PXi="7104
 \mathchardef\PLambda="7103
 \mathchardef\PSigma="7106
 \mathchardef\POmega="710A
 \mathchardef\PUpsilon="7107
                  
 \def\PB      {\ensuremath{B}\xspace}                 
                  
 \def\PD      {\ensuremath{D}\xspace}

 \def\PJ      {\ensuremath{J}\xspace}                 
 \def\PK      {\ensuremath{K}\xspace}

 \def\Pb      {\ensuremath{b}\xspace}                 
 \def\Pc      {\ensuremath{c}\xspace}                 
 \def\Pd      {\ensuremath{d}\xspace}

 \def\Pi      {\ensuremath{i}\xspace}

 \def\Ps      {\ensuremath{s}\xspace}                 
                  
 \def\Pu      {\ensuremath{u}\xspace}

}

\makeatletter
\ifcase \@ptsize \relax%
  \newcommand{\miniscule}{\@setfontsize\miniscule{4}{5}} 
\or
  \newcommand{\miniscule}{\@setfontsize\miniscule{5}{6}}
\or
  \newcommand{\miniscule}{\@setfontsize\miniscule{5}{6}}
\fi
\makeatother

\DeclareRobustCommand{\optbar}[1]{\shortstack{{\miniscule (\rule[.5ex]{1.25em}{.18mm})}
  \\ [-.7ex] $#1$}}

\def\mup        {{\ensuremath{\Pmu^+}}\xspace}
 
\def\mumu       {{\ensuremath{\Pmu^+\Pmu^-}}\xspace}

\def\uquark    {{\ensuremath{\Pu}}\xspace}

\def\dquark    {{\ensuremath{\Pd}}\xspace}
\def\dquarkbar {{\ensuremath{\overline \dquark}}\xspace}

\def\squark    {{\ensuremath{\Ps}}\xspace}
\def\squarkbar {{\ensuremath{\overline \squark}}\xspace}

\def\cquark    {{\ensuremath{\Pc}}\xspace}
\def\cquarkbar {{\ensuremath{\overline \cquark}}\xspace}

\def\bquark    {{\ensuremath{\Pb}}\xspace}
\def\bquarkbar {{\ensuremath{\overline \bquark}}\xspace}

\def\pion   {{\ensuremath{\Ppi}}\xspace}
\def\piz    {{\ensuremath{\pion^0}}\xspace}

\def\pip    {{\ensuremath{\pion^+}}\xspace}

\def\kaon    {{\ensuremath{\PK}}\xspace}
  \def\Kbar    {{\kern 0.2em\overline{\kern -0.2em \PK}{}}\xspace}

\def\KorKbar    {\kern 0.18em\optbar{\kern -0.18em K}{}\xspace}

\def\Kp      {{\ensuremath{\kaon^+}}\xspace}
\def\Km      {{\ensuremath{\kaon^-}}\xspace}

  \def\Dbar    {{\kern 0.2em\overline{\kern -0.2em \PD}{}}\xspace}
\def\D       {{\ensuremath{\PD}}\xspace}

\def\DorDbar    {\kern 0.18em\optbar{\kern -0.18em D}{}\xspace}
\def\Dz      {{\ensuremath{\D^0}}\xspace}

\def\B       {{\ensuremath{\PB}}\xspace}
\def\Bbar    {{\ensuremath{\kern 0.18em\overline{\kern -0.18em \PB}{}}}\xspace}

\def\BorBbar    {\kern 0.18em\optbar{\kern -0.18em B}{}\xspace}

\def\Bc      {{\ensuremath{\B_\cquark^+}}\xspace}
\def\Bcp     {{\ensuremath{\B_\cquark^+}}\xspace}

\def\jpsi     {{\ensuremath{{\PJ\mskip -3mu/\mskip -2mu\Ppsi\mskip 2mu}}}\xspace}

  \def\Y#1S{\ensuremath{\PUpsilon{(#1S)}}\xspace}

\def\Lbar        {{\ensuremath{\kern 0.1em\overline{\kern -0.1em\PLambda}}}\xspace}
\def\LorLbar    {\kern 0.18em\optbar{\kern -0.18em \PLambda}{}\xspace}

\def\BF         {{\ensuremath{\mathcal{B}}}\xspace}

\def\to                 {\ensuremath{\rightarrow}\xspace}

\def\AT#1     {\ensuremath{A_{\mathrm{T}}^{#1}}\xspace}           

\def\C#1      {\ensuremath{\mathcal{C}_{#1}}\xspace}                       
\def\Cp#1     {\ensuremath{\mathcal{C}_{#1}^{'}}\xspace}                   
\def\Ceff#1   {\ensuremath{\mathcal{C}_{#1}^{\mathrm{(eff)}}}\xspace}        
\def\Cpeff#1  {\ensuremath{\mathcal{C}_{#1}^{'\mathrm{(eff)}}}\xspace}     
\def\Ope#1    {\ensuremath{\mathcal{O}_{#1}}\xspace}                       
\def\Opep#1   {\ensuremath{\mathcal{O}_{#1}^{'}}\xspace}

\newcommand{\tev}{\ifthenelse{\boolean{inbibliography}}{\ensuremath{~T\kern -0.05em eV}\xspace}{\ensuremath{\mathrm{\,Te\kern -0.1em V}}}\xspace}
\newcommand{\gev}{\ensuremath{\mathrm{\,Ge\kern -0.1em V}}\xspace}
\newcommand{\mev}{\ensuremath{\mathrm{\,Me\kern -0.1em V}}\xspace}
\newcommand{\kev}{\ensuremath{\mathrm{\,ke\kern -0.1em V}}\xspace}
\newcommand{\ev}{\ensuremath{\mathrm{\,e\kern -0.1em V}}\xspace}
\newcommand{\gevc}{\ensuremath{{\mathrm{\,Ge\kern -0.1em V\!/}c}}\xspace}
\newcommand{\mevc}{\ensuremath{{\mathrm{\,Me\kern -0.1em V\!/}c}}\xspace}
\newcommand{\gevcc}{\ensuremath{{\mathrm{\,Ge\kern -0.1em V\!/}c^2}}\xspace}
\newcommand{\gevgevcccc}{\ensuremath{{\mathrm{\,Ge\kern -0.1em V^2\!/}c^4}}\xspace}
\newcommand{\mevcc}{\ensuremath{{\mathrm{\,Me\kern -0.1em V\!/}c^2}}\xspace}

\def\mum  {\ensuremath{{\,\upmu\mathrm{m}}}\xspace}

\def\invfb   {\ensuremath{\mbox{\,fb}^{-1}}\xspace}

\newcommand{\chisq}{\ensuremath{\chi^2}\xspace}

\def\gsim{{~\raise.15em\hbox{$>$}\kern-.85em
          \lower.35em\hbox{$\sim$}~}\xspace}
\def\lsim{{~\raise.15em\hbox{$<$}\kern-.85em
          \lower.35em\hbox{$\sim$}~}\xspace}

\def\pt         {\mbox{$p_{\mathrm{ T}}$}\xspace}

\def\bcvegpy    {\mbox{\textsc{Bcvegpy}}\xspace}

\def\evtgen     {\mbox{\textsc{EvtGen}}\xspace}

\def\geant      {\mbox{\textsc{Geant4}}\xspace}

\def\photos     {\mbox{\textsc{Photos}}\xspace}

\def\pythia     {\mbox{\textsc{Pythia}}\xspace}

\def\tell1  {TELL1\xspace}
\def\ukl1   {UKL1\xspace}

\usepackage{cite} 
\usepackage{mciteplus}

\def\rk {\ensuremath{{R}_{K/\pi}}\xspace}

\def\BcJpsiK {\ensuremath{\Bcp\to\jpsi\Kp}\xspace}
\def\BcJpsipi {\ensuremath{\Bcp\to\jpsi\pip}\xspace}

\newcommand{\xx}{\ensuremath{\kern 0.5em }}

\def\pp {\ensuremath{pp}\xspace}

\usepackage{multirow} 
\usepackage{booktabs} 
\usepackage{rotating}

\usepackage{longtable}

\begin{document}

\renewcommand{\thefootnote}{\fnsymbol{footnote}}
\setcounter{footnote}{1}

\begin{titlepage}
\pagenumbering{roman}

\vspace*{-1.5cm}
\centerline{\large EUROPEAN ORGANIZATION FOR NUCLEAR RESEARCH (CERN)}
\vspace*{1.5cm}
\noindent
\begin{tabular*}{\linewidth}{lc@{\extracolsep{\fill}}r@{\extracolsep{0pt}}}
\ifthenelse{\boolean{pdflatex}}
{\vspace*{-2.7cm}\mbox{\!\!\!\includegraphics[width=.14\textwidth]{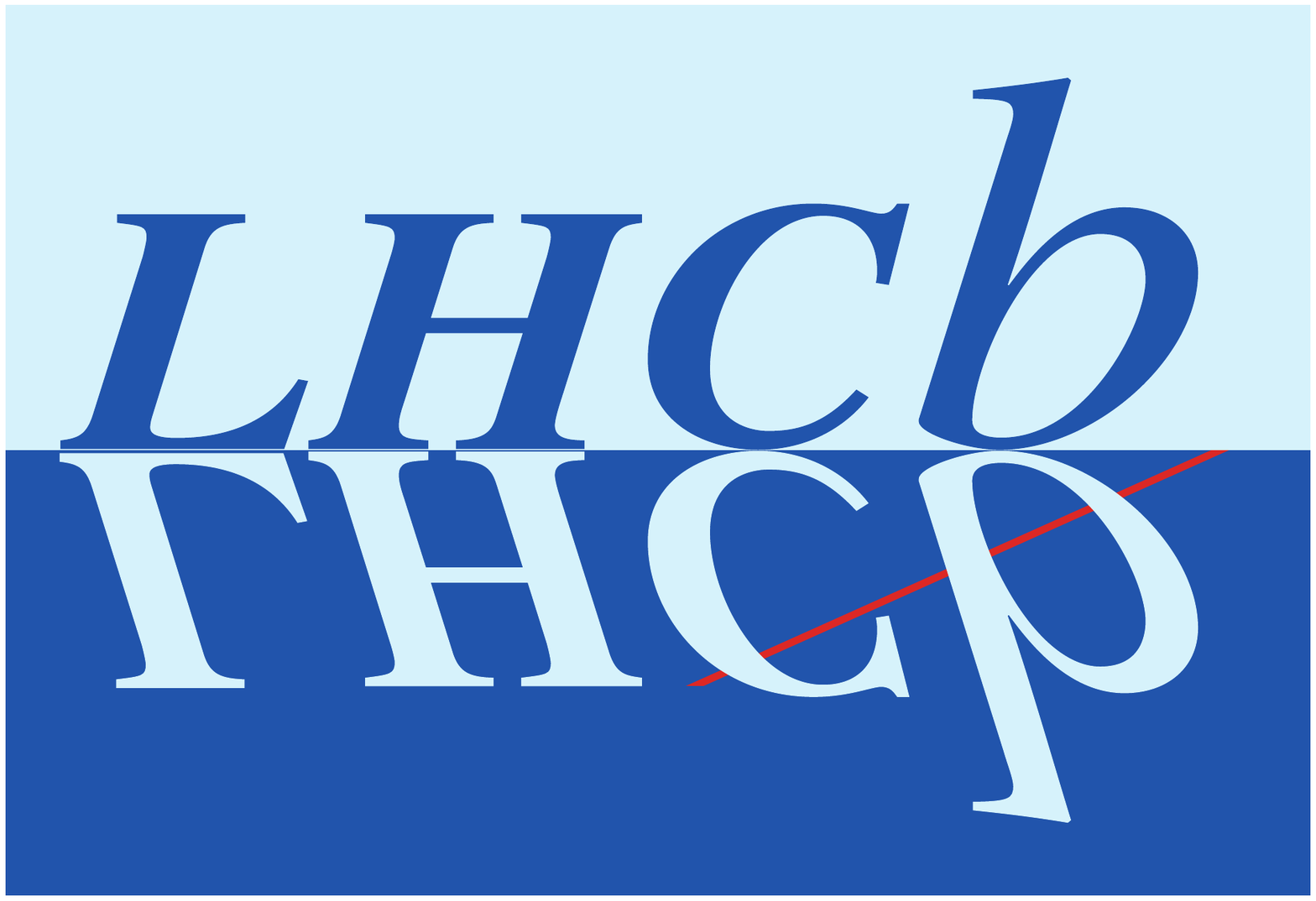}} & &}
{\vspace*{-1.2cm}\mbox{\!\!\!\includegraphics[width=.12\textwidth]{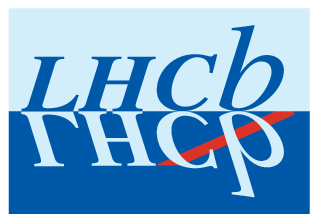}} & &}
\\
 & & CERN-EP-2016-173 \\
 & & LHCb-PAPER-2016-020 \\
 & & \today \\
 & & \\
\end{tabular*}

\vspace*{4.0cm}

{\normalfont\bfseries\boldmath\huge
\begin{center}
  Measurement of the ratio of branching fractions $\BF(\BcJpsiK)/\BF(\BcJpsipi)$
\end{center}
}

\vspace*{2.0cm}

\begin{center}
The LHCb collaboration\footnote{Authors are listed at the end of this paper.}
\end{center}

\vspace{\fill}

\begin{abstract}
  \noindent
  The ratio of branching fractions $\rk \equiv \BF(\BcJpsiK)/\BF(\BcJpsipi)$
  is measured with \pp collision data collected by the LHCb experiment
  at centre-of-mass energies of $7\tev$ and $8\tev$,
  corresponding to an integrated luminosity of 3\invfb.
  It is found to be $ \rk = 0.079\pm0.007\pm0.003$,
  where the first uncertainty is statistical and the second is systematic.
  This measurement is consistent with the previous LHCb result,
  while the uncertainties are significantly reduced.
\end{abstract}

\vspace*{2.0cm}

\begin{center}
  Published in JHEP 09(2016)153
\end{center}

\vspace{\fill}

{\footnotesize
\centerline{\copyright~CERN on behalf of the \lhcb collaboration, licence \href{http://creativecommons.org/licenses/by/4.0/}{CC-BY-4.0}.}}
\vspace*{2mm}

\end{titlepage}

\newpage
\setcounter{page}{2}
\mbox{~}

\cleardoublepage

\renewcommand{\thefootnote}{\arabic{footnote}}
\setcounter{footnote}{0}

\pagestyle{plain}
\setcounter{page}{1}
\pagenumbering{arabic}

\section{Introduction}
\label{sec:Introduction}

The \Bc meson,
the lightest $\bar{b} c$ bound state, can only decay weakly.
Since it contains only heavy quarks,
its decays can be analysed 
using various theoretical approaches, including 
QCD-based methods~\cite{Kiselev:2000pp, Jones:1998ub, Kiselev:1999sc} and 
QCD-inspired phenomenological models~\cite{Eichten:1994gt,Naimuddin:2012dy}.
A measurement of the weak decay properties of $\Bc$ mesons can test these approaches and
provide  insight into the dynamics of the heavy quarks in the $\Bc$ meson.

The exclusive decay\footnote{The inclusion of charge-conjugate processes is implied throughout the paper.} $\BcJpsiK$ is of particular interest since it 
proceeds via a $\bquarkbar\to\cquarkbar\uquark\squarkbar$ transition and thus is CKM-suppressed 
by a factor $|V_{us}/V_{ud}|^2\sim0.05$ with respect to $\BcJpsipi$,
where the dominant amplitude is a $\bquarkbar\to\cquarkbar\uquark\dquarkbar$ transition. 
In addition to the CKM matrix elements, the ratio of branching fractions
$\rk \equiv \BF(\BcJpsiK)/\BF(\BcJpsipi)$ depends on the form factors of the two decays.
Theoretical calculations of $\rk$ have been carried out
using approaches that handle the non-factorisable and non-perturbative
contributions in different ways, yielding values in the range 
from $0.05$ to $0.10$~\cite{Chang:1992pt, Liu:1997hr,Anisimov:1998uk,
Colangelo:1999zn, AbdElHady:1999xh, Kiselev:2000pp, Kiselev:2002vz, Ebert:2003cn,
Ivanov:2006ni, Naimuddin:2012dy, Qiao:2012hp,Ke:2013yka}.

The decay $\BcJpsiK$ was first observed by the \lhcb collaboration, 
which  reported a measurement of $\rk=0.069\pm0.019\pm0.005$~\cite{LHCb-PAPER-2013-021}. 
The uncertainty on this value is too large to discriminate between the predictions quoted above. 
The \pp  data sample used in Ref. \cite{LHCb-PAPER-2013-021}, taken at a centre-of-mass 
energy of $7\tev$ and corresponding to an integrated luminosity of $1\invfb$, 
is now reanalysed in this paper
together with an additional sample taken at a centre-of-mass energy 
of $8\tev$ and corresponding to an integrated luminosity of $2\invfb$.
Owing to improvements in the analysis method as well as the increase in the data sample size, 
the statistical uncertainty is reduced by a factor of more than two.
The systematic uncertainty is also reduced.

\section{Detector and simulation}
\label{sec:Detector}
The \lhcb detector~\cite{Alves:2008zz,LHCb-DP-2014-002} is a single-arm forward
spectrometer covering the \mbox{pseudorapidity} range $2<\eta <5$,
designed for the study of particles containing \bquark or \cquark
quarks. The detector includes a high-precision tracking system
consisting of a silicon-strip vertex detector surrounding the $pp$
interaction region, a large-area silicon-strip detector located
upstream of a dipole magnet with a bending power of about
$4{\mathrm{\,Tm}}$, and three stations of silicon-strip detectors and straw
drift tubes placed downstream of the magnet.
The tracking system provides a measurement of charged particle momentum, $p$, with
a relative uncertainty that varies from 0.5\% at low momentum to 1.0\% at 200\gevc.
The minimum distance of a track to a primary vertex (PV), the impact parameter, is measured with a resolution of $(15+29/\pt)\mum$,
where \pt (in GeV/$c$) is the component of the momentum transverse to the beam direction.
Different types of charged hadrons are distinguished using information
from two ring-imaging Cherenkov (\rich) detectors. 
Photons, electrons and hadrons are identified by a calorimeter system consisting of
scintillating-pad and preshower detectors, an electromagnetic
calorimeter and a hadronic calorimeter. 
Muons are identified by a
system composed of alternating layers of iron and multiwire
proportional chambers.

The trigger comprises a hardware stage and a software stage.
The hardware trigger employed in this analysis uses information 
from the muon system to select single muons or muon pairs, 
applying $\pt$ requirements.
The subsequent software trigger is composed of two stages, the first of which 
performs a partial reconstruction and requires either a pair of  well-reconstructed, 
oppositely charged muons having an invariant mass above $2.7 \gevcc$, or a
single well-reconstructed muon. The second stage of the software trigger applies
a full event reconstruction, and requires at least one of the following two
conditions to be fulfilled: either two opposite-sign muons must form a
good-quality vertex that is well separated from all of the primary vertices and
must have an invariant mass within $120\mevcc$ of the known $\jpsi$ mass
\cite{PDG2014}, or an algorithm using a boosted decision tree  must identify a
two- or three-track vertex that is well separated from all of the primary
vertices and includes a muon among the constituent tracks.
The same trigger requirements are used to select both $\Bc\to\jpsi\Kp$ and
$\Bc\to\jpsi\pip$ decays, due to the similarity in their kinematic distributions.

In the simulation, $pp$ collisions are generated using \pythia
6~\cite{Sjostrand:2006za} with a specific \lhcb
configuration~\cite{LHCb-PROC-2010-056}, or, for the hard process
$gg\to\Bc+b+\cquarkbar$ that is the dominant source of $\Bc$ mesons, using the
dedicated generator \bcvegpy~\cite{Chang:2003cq,Chang:2005hq}.  Decays of
hadronic particles are described by \evtgen~\cite{Lange:2001uf}, in which
final-state radiation is generated using \photos~\cite{Golonka:2005pn}.  The
interaction of the generated particles with the detector, and its response, are
implemented using the \geant toolkit~\cite{Allison:2006ve, *Agostinelli:2002hh}
as described in Ref.~\cite{LHCb-PROC-2011-006}.

\section{Event selection}
Candidate $\Bc \to \jpsi\,h^+$ decays, 
with $\jpsi \to \mumu$ and $h^+$ being a $K^+$ or $\pi^+$,
are reconstructed as follows. 
First a loose preselection is applied. Pairs of oppositely charged, well-reconstructed
muon tracks with $\pt>550\mevc$ consistent with originating from a 
common vertex are combined to form $\jpsi\to\mumu$ candidates.
Hadron ($h^+$) candidates are selected from well-reconstructed tracks with $\pt>500\mevc$, inconsistent with originating from any PV and with the muon hypothesis.
Candidate $\Bc\to\jpsi\Kp$ and $\Bc\to\jpsi\pip$ decays are formed from $\jpsi h^+$ combinations 
that originate from a common vertex. They must also be within $500\mevcc$ of the known $\Bc$ mass \cite{PDG2014}. 
The impact parameter $\chisq$, $\chisq_{\rm IP}$, which is defined as the difference 
in the vertex fit $\chisq$ of the PV with and without the particle 
under consideration, is required to be less than 16 for the $\Bc$ candidates.

A multivariate classifier using a boosted decision tree (BDT) \cite{Breiman} is constructed to further suppress the combinatorial background.
The kinematic variables used as inputs to the BDT are chosen to discriminate between signal and background. 
The twelve variables chosen are:  the $\chisq_{\rm IP}$ of the \Bc, \jpsi, \mup, $\mu^-$ and $h^+$ candidates;
the \pt of the \jpsi, \mup, $\mu^-$ and $h^+$ candidates;
the $\chisq$ per degree of freedom of the $\Bc$ vertex fit; and the decay time and the decay length of the $\Bc$ candidate.
Since the kaon-pion mass difference is small compared with the energy release of 
$\Bc \to \jpsi h^+$ decays, the distributions of the BDT variables
are similar for $\Bc \to \jpsi K^+$ and $\Bc \to \jpsi \pi^+$ decays.
The BDT is trained with simulated $\Bc \to \jpsi \pi^+$ decays to represent 
both the $\Bc \to \jpsi K^+$  and the $\Bc \to \jpsi \pi^+$  signals, 
and with events from the upper mass sideband of the $\Bc\to\jpsi\pip$ candidates in data, 
$[6444,~6528]\mevcc$, to represent the combinatorial background.
For one third of the events in the training samples the centre-of-mass energy is $7\tev$, 
and for the rest it is $8\tev$ in accordance with the ratio of integrated luminosities.
Since the BDT does not use any particle identification information,
it  selects both $\Bc \to \jpsi \Kp$ and $\Bc \to \jpsi \pip$ candidates.
Particle identification requirements using information from the \rich subdetectors
are then applied to the hadrons
to obtain two mutually exclusive samples of $\Bc \to \jpsi \Kp$ and $\Bc \to \jpsi \pip$ candidates.

The BDT and particle identification requirements
are optimised sequentially on the sample of $\Bc\to\jpsi\Kp$ candidates that
pass the loose preselection to maximise $N_K/\sqrt{N_{\rm tot}}$, 
where $N_{\rm tot}$ is the total number of candidates 
within $\pm 3$ times the mass resolution around the known $\Bc$ mass. 
Here $N_K$ refers to the $\Bc \to \jpsi \Kp$ signal yield and 
is estimated to be $(N_{\rm tot}-N_{\rm comb})/(1+1/(r_{\rm eff}\rk))$, 
where the value of $\rk$ is taken from the previous LHCb measurement~\cite{LHCb-PAPER-2013-021},
$N_{\rm comb}$ is the number of combinatorial background events in the signal region
extrapolated from the upper sideband, 
and $r_{\rm eff}$ represents the ratio of  the numbers of $\Bc \to \jpsi \Kp$ and $\Bc \to \jpsi \pi^+$
events that pass the $\Bc \to \jpsi \Kp$  selection and fall in the signal window.
After this optimisation, the BDT rejects more than $99.8\%$ of 
the combinatorial background and keeps around $70\%$ of  $\Bc \to \jpsi h^+$ events.
This particle identification requirement has an efficiency of about 
$70\%$ for $\Bc \to \jpsi \Kp$ and $87\%$ for $\Bc \to \jpsi \pip$, while the
probabilities for a charged kaon to be misidentified as a pion and a charged
pion to be misidentified as a kaon are below $7\%$ and $1\%$, respectively.

\section{Signal yields and efficiency correction}
\label{sec:yields}
The measurement is made by evaluating
\begin{equation} \label{eq:Rk}
 \rk \equiv\frac{\BF(\BcJpsiK)}{\BF(\BcJpsipi)}
     = \frac{N(\BcJpsiK)}{N(\BcJpsipi)} \times
       \frac{\epsilon(\BcJpsipi)}{\epsilon(\BcJpsiK)},
\end{equation}
where  $N(\BcJpsiK)$ and $N(\BcJpsipi)$ are the signal yields,
and $\epsilon(\BcJpsiK)$ and $\epsilon(\BcJpsipi)$ are the total efficiencies
estimated with simulation and control samples of data.

The signal yields $N(\BcJpsiK)$ and $N(\BcJpsipi)$
 are obtained from a simultaneous unbinned maximum likelihood fit to the distribution of
\Bc candidate masses  in the range 6000 to 6600 \mevcc.
These candidates include the part of the background training sample that passes the full selection;
the effect of doing so has been investigated and found not to lead to any systematic bias.
The fit model includes
components due to signal, combinatorial background
and misidentified decays (\BcJpsipi misidentified as \BcJpsiK, or vice versa).

A partially reconstructed  background component is included for \BcJpsipi.
 This background is mainly due to  $\Bc \to \jpsi \rho^+$ decays
followed by $\rho^+ \to \pip \piz$.
The data show no clear indication of  partially reconstructed  background
for  \BcJpsiK. A systematic uncertainty is assigned due to the non-inclusion of this background component.

The signal mass distribution of $\Bc\to\jpsi\,h^+$ is described by the
sum of two double-sided Crystal Ball ($F^{\rm DSCB}$) functions consisting
of a Gaussian core and power law tails on both sides,
\begin{eqnarray}
f_{\rm sig}(M_{\Bc})=\alpha F^{\rm DSCB}_1(M_{\Bc}) + (1-\alpha)F^{\rm DSCB}_2(M_{\Bc}),
\end{eqnarray}
where $M_{\Bc}$ is the invariant mass of the $\mumu\,h^+$ combination with the
mass of the $\mumu$ pair constrained to the known $\jpsi$ mass.
In the simultaneous fit, the Gaussian mean and the core mass resolution
$\sigma_1$ of $F^{\rm DSCB}_1$ are allowed to vary, and set to be the same
for both the $\Bc\to\jpsi\pip$ and the $\Bc\to\jpsi\Kp$ decays.
The tail parameters, the fraction $\alpha$ and the ratio $\sigma_2/\sigma_1$
of the core-mass resolutions of $F^{\rm DSCB}_1$ and $F^{\rm DSCB}_2$
are fixed to the values  obtained in simulation.

The combinatorial background for each decay mode is modelled by an exponential distribution.
Background arising from misidentified $\Bc \to J/\psi h^+$ decays
is described by a DSCB function, with
 shape and mass offset relative to the signal peak  derived from simulation for each mode separately.
The invariant mass distribution of the partially reconstructed background is
taken to be an ARGUS function\cite{Albrecht:1990am} convolved with a Gaussian resolution function.
The mean and the width parameters of the resolution
 function are set to be zero and $\sqrt{\alpha\sigma_1^2+(1-\alpha)\sigma_2^2}$.

The parameters estimated from the simultaneous fit are:
the yield $N(\BcJpsipi)$,  the yield ratio $N(\BcJpsiK)/N(\BcJpsipi)$,
the numbers of combinatorial background events for  \BcJpsiK and \BcJpsipi decays,
the number of misidentification  background events for each of the decay modes,
the number of partially reconstructed background events for the \BcJpsipi decay,
and the shape parameters describing the signal and background distributions.

\begin{figure}[tb]
  \begin{center}
    \includegraphics[width=0.48\linewidth]{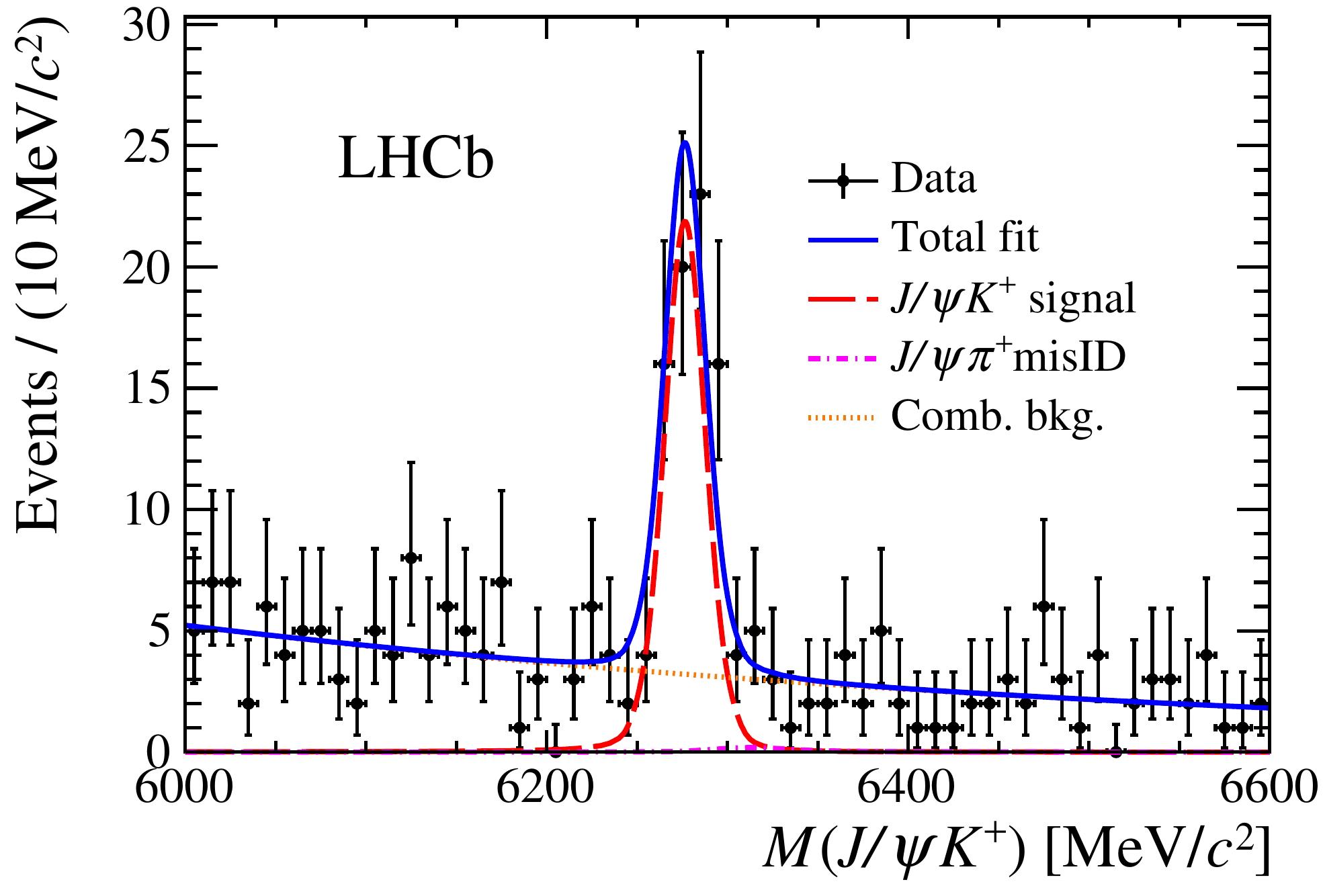}
    \includegraphics[width=0.48\linewidth]{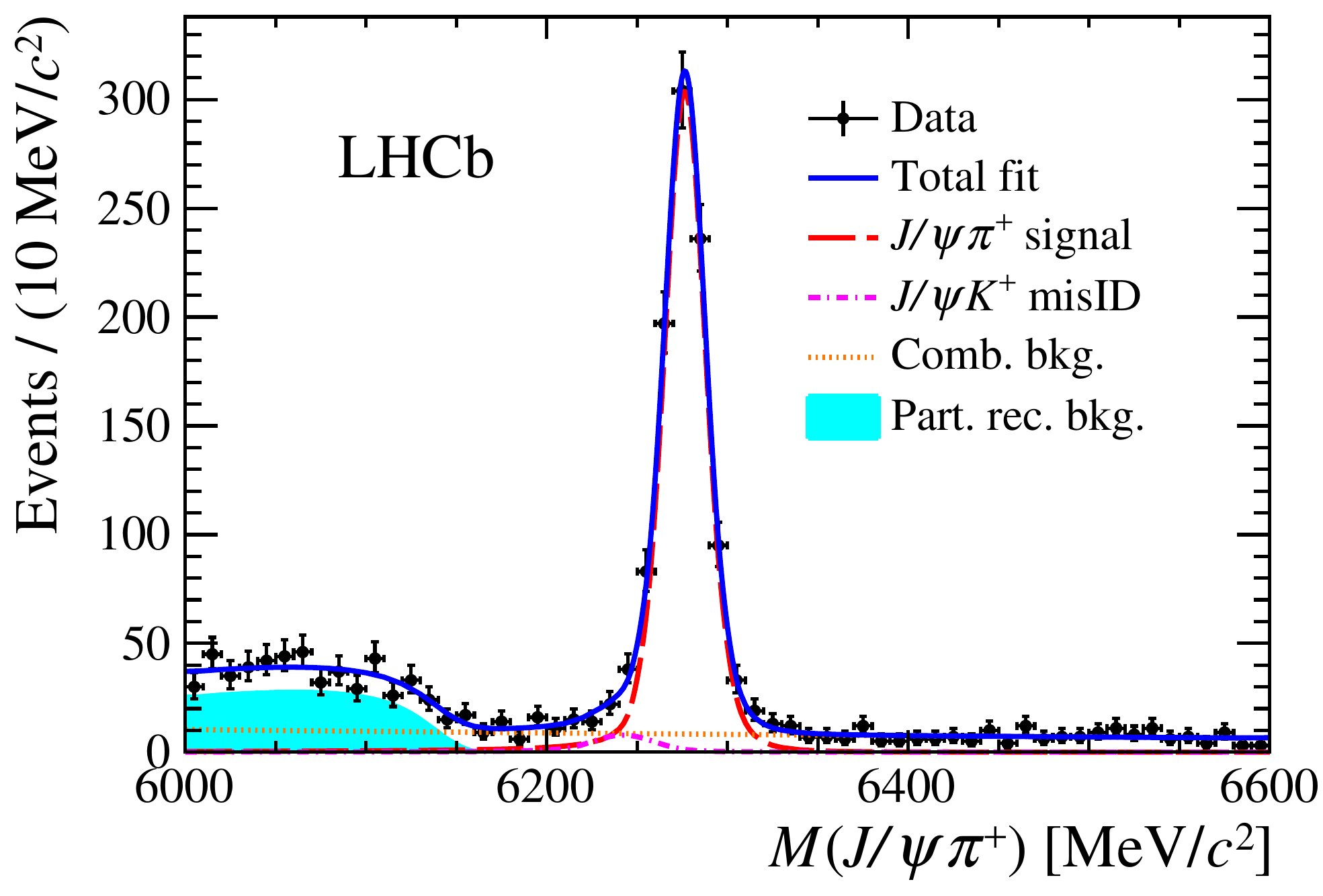}\\
    \vspace*{0.5cm}
    \includegraphics[width=0.48\linewidth]{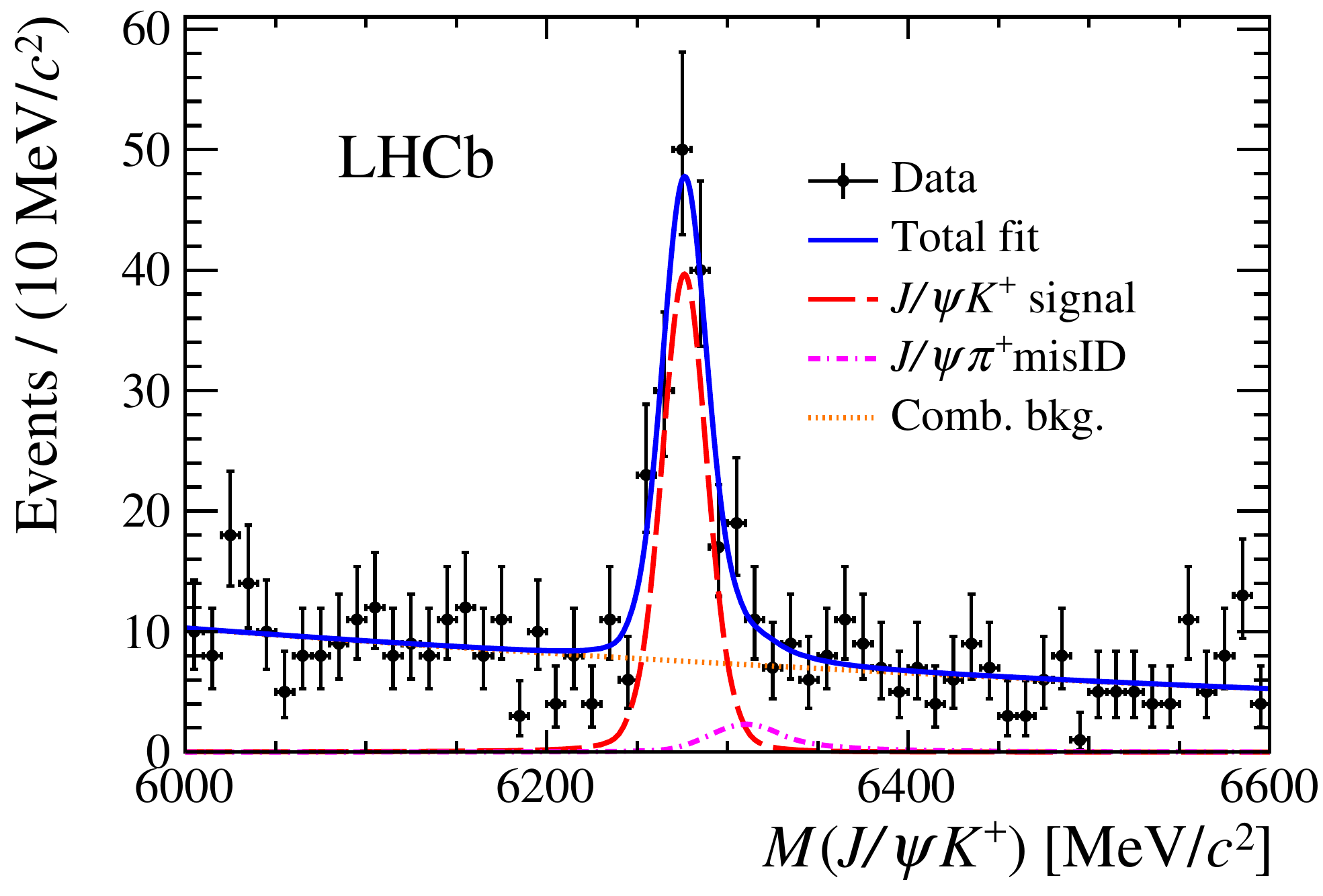}
    \includegraphics[width=0.48\linewidth]{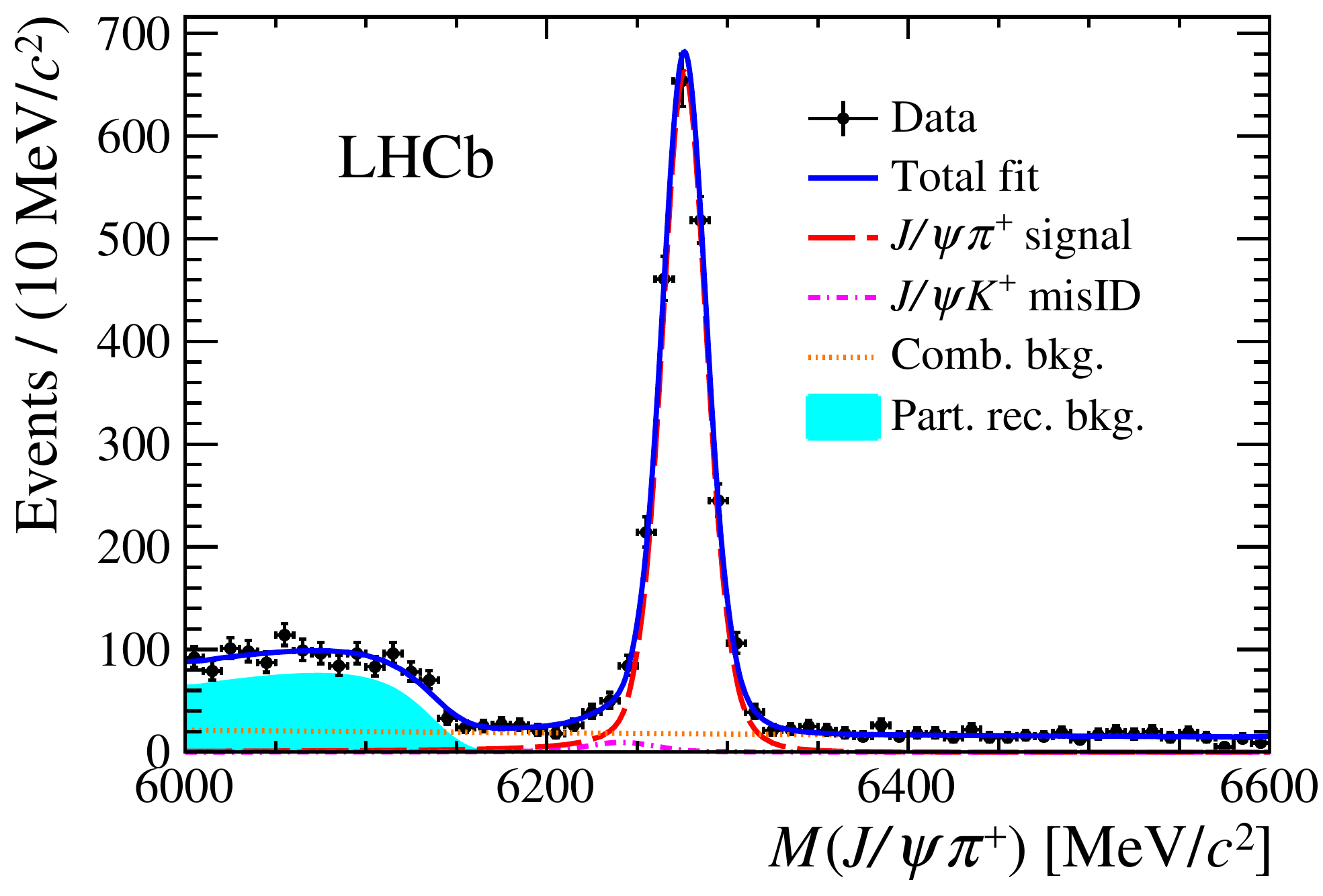}
    \vspace*{-0.5cm}
  \end{center}
  \caption{    Fits to the reconstructed \BcJpsiK (left) and \BcJpsipi (right) mass distributions
           using 7 TeV (top) and 8 TeV  (bottom) data samples.
The contributions from the signal, the misidentification background,
the combinatorial background and the partially reconstructed background
are indicated in the figures.
           }
  \label{fig:fit}
\end{figure}

The results of the separate fits to the 7 and 8\tev samples are shown in Fig.~\ref{fig:fit}.
In the 7\tev sample, the yield $N(\Bc \to \jpsi \pip)$ is found  to be $954 \pm 36$ and the yield ratio
$N(\BcJpsiK)/N(\BcJpsipi)$  is found
to be $0.069 \pm 0.010$. The corresponding values in the 8\tev sample are $2253 \pm 53$ and $0.059 \pm 0.006$.

The ratio of branching fractions $\rk$
is obtained by correcting the yield ratio
with the relative efficiency, as shown in Eq.~\ref{eq:Rk}.
The total efficiencies include contributions from
the LHCb detector acceptance and from selection, trigger and particle identification requirements.
The selection and trigger efficiencies are calculated from simulated samples.
The simulated events are weighted to account for
differences from data in the track multiplicity distribution.
It has been checked that after this weighting, the distributions of the variables used as inputs to the BDT
are similar in data and simulation.
The particle identification efficiencies for hadrons are evaluated from simulation calibrated with
a control sample of $D^{*+} \to \Dz \pip$, $\Dz \to \Km \pip$ decays.
The efficiency ratio is determined to be
$\epsilon(\BcJpsipi)/ \epsilon(\BcJpsiK) =  1.277\pm 0.007$
and $1.284 \pm 0.006$ for $7 \tev$  and $8 \tev$ data, respectively.
The  efficiency difference between \BcJpsipi and \BcJpsiK mainly arises from
particle identification for the hadrons.

\section{Systematic uncertainties}
Since the running conditions changed between $7\tev$ and $8\tev$, 
the systematic uncertainties on $\rk$ are determined separately for the two samples. 
Table~\ref{tab:sys} summarises the relative systematic uncertainties associated with the mass fit and efficiency estimates
that affect the ratio of branching fractions. The sources of these uncertainties are discussed below.

Each of the systematic uncertainties associated with
 the mass fit is studied by generating an ensemble of pseudoexperiments according to the nominal model 
 described above
and fitting them with an alternative model. The difference in the mean values of $\rk$ obtained 
is taken as the systematic uncertainty.  

Changing the signal model from
the sum of two DSCB functions to a single DSCB function leads to 
relative systematic uncertainties of $0.5\%$ and $0.8\%$ for the 7\tev and 8\tev data, respectively.
Using a third-order polynomial in place of an exponential function for the combinatorial background 
changes the mean values of $\rk$ by $1.1\%$ and $0.5\%$ for the two samples.

In the nominal fit,  the partially reconstructed background
 is neglected for \BcJpsiK decays for reasons of fit stability.
 The associated systematic uncertainties are estimated 
by including such a component in the same way as was done for \BcJpsipi decays,
and are found to be $3.3\%$ and $3.2\%$ for the 7 and $8\tev$ data, respectively.
Using the sum of two DSCB functions instead of a single DSCB function for the misidentification
background events changes the mean values of $\rk$ by $0.2\%$ and $0.0\%$ for the two samples. 

The selection and trigger efficiencies are calculated with simulated samples. 
Systematic effects on the efficiency evaluation due to differences between data and simulation 
in the distributions of variables such as muon momentum and \Bc decay time  are investigated. 
Such effects are found to cancel in the efficiency ratio and 
thus have negligible impact on $\rk$.

The kaon and pion  identification efficiencies are measured as functions of momentum and pseudorapidity
with a control sample of $D^{*+} \to \Dz \pip$, $\Dz \to \Km \pip$ decays,
and represented by two-dimensional histograms. 
When the histogram binning is varied, the largest  changes in the efficiency ratio seen 
are $0.2\%$ and $0.1\%$ for the $7\tev$ and 8\tev samples, 
and these values are assigned as the corresponding relative systematic uncertainties.

The simulation accounts for the different interaction cross-sections
of pions and kaons with matter. However, if the amount of material in 
the detector is not modelled correctly, 
this would alter the efficiency ratio. A systematic uncertainty of 
$0.3\%$ associated with this effect is assigned for both $7\tev$ and $8 \tev$ samples.
Adding all of the above contributions in quadrature, the total relative systematic uncertainties 
on $\rk$ are $3.5\%$ and $3.4\%$ for the $7\tev$ and $8\tev$ results.

\begin{table}[h]
 \caption{Summary of the relative systematic uncertainties on $\rk$.}
\vspace*{-0.5cm}
 \begin{center}\begin{tabular}{lcc}
 \hline
 & $7\tev$ & $8\tev$ \\
 \hline
 Signal model & $0.5\%$ & $0.8\%$\\
 Combinatorial background & $1.1\%$ & $0.5\%$\\
 Partially reconstructed background & $3.3\%$ & $3.2\%$\\
 Misidentification background  & $0.2\%$ & $0.0\%$\\
 Particle identification efficiency & $0.2\%$ & $0.1\%$\\
 Detector material & $0.3\%$ & $0.3\%$\\
 \hline
 Total & $3.5\%$ & $3.4\%$\\
 \hline
 \end{tabular}\end{center}
 \label{tab:sys}
 \end{table}

\section{Results and summary}
\label{sec:Results}
Using the yield and efficiency ratios,
the ratio of branching fractions of \BcJpsiK and \BcJpsipi is evaluated as
\begin{equation*}
\rk= 0.089\pm0.013\pm0.003
\end{equation*}
for the 7\tev data sample 
and 
\begin{equation*}
\rk = 0.075\pm0.008\pm0.003
\end{equation*}
 for the 8\tev sample,
where the first uncertainties  are statistical and the second are systematic.

The two results are combined by evaluating their weighted average. 
The systematic uncertainties of both measurements are dominated by 
the contribution from the non-inclusion of the partially reconstructed 
background for $\Bc\to\jpsi\Kp$ decays, and so are assumed to be fully 
correlated, while their statistical uncertainties are independent.
The combined measurement for the $7\tev$ and $8\tev$ data sample is   
\begin{equation*}
\rk = 0.079\pm0.007\pm0.003\,.
\end{equation*}
This is consistent with the previous LHCb measurement $\rk = 0.069\pm0.019\pm0.005$~\cite{LHCb-PAPER-2013-021}, which was based on the 7\tev data alone. 
The uncertainties are significantly reduced due to both the increased sample size and the improved event selection.
The result supersedes the previous measurement \cite{LHCb-PAPER-2013-021} 
and agrees with  the theoretical predictions in Refs. \cite{
Kiselev:2000pp, Chang:1992pt, Liu:1997hr,
AbdElHady:1999xh, Ebert:2003cn,
Ivanov:2006ni, Naimuddin:2012dy, Qiao:2012hp, Ke:2013yka}, but disfavours that based on QCD sum rules \cite{Kiselev:2002vz}.

\section*{Acknowledgements}

\noindent We express our gratitude to our colleagues in the CERN
accelerator departments for the excellent performance of the LHC. We
thank the technical and administrative staff at the LHCb
institutes. We acknowledge support from CERN and from the national
agencies: CAPES, CNPq, FAPERJ and FINEP (Brazil); NSFC (China);
CNRS/IN2P3 (France); BMBF, DFG and MPG (Germany); INFN (Italy); 
FOM and NWO (The Netherlands); MNiSW and NCN (Poland); MEN/IFA (Romania); 
MinES and FANO (Russia); MinECo (Spain); SNSF and SER (Switzerland); 
NASU (Ukraine); STFC (United Kingdom); NSF (USA).
We acknowledge the computing resources that are provided by CERN, IN2P3 (France), KIT and DESY (Germany), INFN (Italy), SURF (The Netherlands), PIC (Spain), GridPP (United Kingdom), RRCKI and Yandex LLC (Russia), CSCS (Switzerland), IFIN-HH (Romania), CBPF (Brazil), PL-GRID (Poland) and OSC (USA). We are indebted to the communities behind the multiple open 
source software packages on which we depend.
Individual groups or members have received support from AvH Foundation (Germany),
EPLANET, Marie Sk\l{}odowska-Curie Actions and ERC (European Union), 
Conseil G\'{e}n\'{e}ral de Haute-Savoie, Labex ENIGMASS and OCEVU, 
R\'{e}gion Auvergne (France), RFBR and Yandex LLC (Russia), GVA, XuntaGal and GENCAT (Spain), Herchel Smith Fund, The Royal Society, Royal Commission for the Exhibition of 1851 and the Leverhulme Trust (United Kingdom).

\addcontentsline{toc}{section}{References}
\setboolean{inbibliography}{true}
\bibliographystyle{LHCb}
\bibliography{main,LHCb-PAPER,LHCb-CONF,LHCb-DP,LHCb-TDR,MyBib}

\newpage

\newpage
\centerline{\large\bf LHCb collaboration}
\begin{flushleft}
\small
R.~Aaij$^{39}$,
B.~Adeva$^{38}$,
M.~Adinolfi$^{47}$,
Z.~Ajaltouni$^{5}$,
S.~Akar$^{6}$,
J.~Albrecht$^{10}$,
F.~Alessio$^{39}$,
M.~Alexander$^{52}$,
S.~Ali$^{42}$,
G.~Alkhazov$^{31}$
% ...

{\footnotesize \it
% ...
$ ^{5}$Clermont Universit{\'e}, Universit{\'e} Blaise Pascal, CNRS/IN2P3, LPC, Clermont-Ferrand, France\\
$ ^{6}$CPPM, Aix-Marseille Universit{\'e}, CNRS/IN2P3, Marseille, France\\
% ...
$ ^{10}$Fakult{\"a}t Physik, Technische Universit{\"a}t Dortmund, Dortmund, Germany\\
% ...
$ ^{31}$Petersburg Nuclear Physics Institute (PNPI), Gatchina, Russia\\
% ...
$ ^{38}$Universidad de Santiago de Compostela, Santiago de Compostela, Spain\\
$ ^{39}$European Organization for Nuclear Research (CERN), Geneva, Switzerland\\
% ...
$ ^{42}$Nikhef National Institute for Subatomic Physics, Amsterdam, The Netherlands\\
% ...
$ ^{47}$H.H. Wills Physics Laboratory, University of Bristol, Bristol, United Kingdom\\
% ...
$ ^{52}$School of Physics and Astronomy, University of Glasgow, Glasgow, United Kingdom\\
% ...
}
\end{flushleft}

\end{document}